# Orbital-Angular-Momentum Entangled Photon Emission from Circular Currents in Semiconductor-Superconductor Structures

Avi Koriat[*], Ankit Kumar[*,†] and Alex Hayat

Department of Electrical Engineering, Technion Israel Institute of Technology, Haifa 3200003, Israel

We theoretically demonstrate that a superconducting circular current induced in a semiconductor results in emission of orbital-angular-momentum (OAM) entangled photon pairs upon carrier recombination. Combining the macroscopic Ginzburg-Landau theory and the microscopic Bardeen-Cooper-Schrieffer (BCS) theory, we investigate the emission of a superconducting light-emitting diode (SLED) with a spatially varying phase profile in the superconducting order parameter. We show that in the active region of the SLED with a circular supercurrent, radiative recombination processes inherit the order parameter phase and result in photon pairs emitted into modes of different OAM quantum numbers. We demonstrate that coherent superposition of superconducting qubit eigenstates can also be mapped onto a coherent superposition of emitted photon states. We also show that other recombination processes due to thermally excited quasi particles do not significantly degrade the state purity. Our results introduce an original scheme for generating OAM-entangled photons enabling a new method of transmitting superconducting qubit information to photonic channels thereby bridging the gap between solid-state and photon-based platforms for quantum communications and information processing.

[*] A. K. and A. K. contributed equally to this work.
[†] Corresponding Author: ankitbhu3258@gmail.com

Superconducting quantum devices are at the heart of various technologies and applications in the field of quantum computation [1,2], and sensing [3,4,5,6]. These devices include, among others, the persistent-current flux qubit – a device that can operate as a two-level system characterized by the orientation of its supercurrent, namely clockwise, counterclockwise and their superpositions [7,8,9,10]. Superconducting qubits have in recent years become the leading candidates for the realization of a large-scale quantum computer [10,11,12,13] due to their compatibility with on-chip radio-frequency manipulation and readout [14,15]. However, while superconducting qubits excel in on-chip processing [10-12], their strong environmental coupling and the need for extremely low temperatures [16,17] limit their use in long-distance quantum communication, teleportation, and distributed quantum networks [18]. Photonic quantum channels operate at room temperature, have extremely weak photon-photon and photon-environment interactions [19,20] exceptionally long dephasing times: tens of microseconds in low-loss optical fibers and theoretically unlimited in free space [21,22,23,24,25]. These properties make photons suitable candidate for applications such as long-distance quantum communication [26,27,28,29], teleportation [30,31], and quantum metrology [32,33] although is very challenging implementing photonic quantum gates for computing.

SLED [34,35,36,37,38,39,40,41,42,43,44,45,46,47,48,49,50] is a promising candidate for creating an interface between solid-state-based and photonic-based quantum technologies, being a superconducting device that emits photon pairs through radiative recombination of Cooper pairs (CP). These hybrid superconductor–semiconductor (super–semi) devices [34-50], enabled by proximity-induced superconducting contacts on semiconductors, have emerged as a key platform for advanced quantum technologies [1-6,10-13,26-33]. Such hybrid platforms enable efficient electrical injection, controlled recombination, and enhanced light–matter interaction, while maintaining compatibility with nanoscale device architectures. As a result, they support a wide range of applications, including light-emitting devices, gate-tunable Josephson junctions, and hybrid quantum circuits [34-50]. Moreover, the condensation of a macroscopic number of CP in the active region of the SLED due to the superconducting proximity effect [51,52] allows for emission at high rates from small volumes [53,54,55,56]. Research in SLEDs has been focused thus far [57,58] on the generation of polarization-entangled photon pairs, exploiting the singlet state of a CP to transfer spin angular momentum (SAM) quantum information from the superconducting condensate to photons. Thus, polarization-entangled photon pairs emitted from SLEDs reflect a local property of a BCS condensate intrinsic to a single CP, and not to the collective behavior of a CP condensate. Consequently, the polarization entanglement of photon pairs emitted from SLEDs cannot convey information about the macroscopic states of a superconducting qubit, which are characterized by a spatial order parameter phase profile with the same orientation as the supercurrent [59]. The transmission of quantum information from these superconducting qubit states onto photons necessitates a coupling mechanism between the spatial distribution of the SC order parameter and that of the electromagnetic field.

Here we theoretically demonstrate a direct superconducting to photon quantum interface. We show that two-photon emission occurs via radiative recombination of CP injected into a p-n quantum-well junction in a superconductor–semiconductor–superconductor structure. The superconducting cathode is shaped as a closed ring with a Josephson junction, while the anode is a disk-shaped superconducting contact. The active semiconductor region inherits the azimuthal phase profile of the superconducting ring, thereby transferring the macroscopic qubit state directly into the photonic OAM degree of freedom. This mechanism enables the generation of photon pairs entangled in discrete OAM modes, providing access to a high-dimensional photonic Hilbert space [60]. This property admits an additional degree of freedom for the encoding of quantum information, making it useful in applications such as quantum computation, quantum communication and quantum memory [61, 62].

We consider a quantum-well SLED in which both the n and p-type regions are contacted by superconductors with distinct geometries: a ring-shaped cathode and a disk-shaped anode (Fig. 1(a)). The cathode forms a flux qubit (superconducting ring with a Josephson junction) whose phase distribution is governed by the competition between the Josephson energy ($E_J$) and inductive energy ($E_L$). In the stiff-junction limit ($E_J \gg E_L$), the superconducting phase resides primarily along the ring, enabling persistent-current qubit operation [7,63]. Figure 1(b) shows the cross-section view, and Fig. 1(c) shows a plot of band diagram across the different layers of the structure, and the superconducting gap is penetrating the semiconductors [64]. The device thus realizes a superconductor–semiconductor–superconductor architecture, with induced superconductivity in both the conduction (CB) and valence bands (VB). The key asymmetry arises from the distinct geometries of the superconducting contacts: a ring-shaped cathode contact (allowing quantized persistent currents with nonzero winding number $l_s$) and a disk-shaped anode contact, consistent with experimentally demonstrated geometries as in Ref. [35]. To achieve e.g., $l_s = 2$ in the cathode condensate while maintaining $l_s = 0$ in the anode, standard flux bias method can be used. The anode, being in contact with a disk-shaped superconductor, remains field-free due to the Meissner effect. The voltage bias (anode superconductor at $V = 0$, while a voltage bias is applied on the cathode superconductor $V \neq 0$) will drive radiative recombination and sets the emission frequency via the AC Josephson effect. Excitation of specific superpositions is achieved using standard techniques flux bias methods [8,65,66]. Readout is performed by the usual SQUID [8,65,67] or lumped-element resonator [66,68,69]. In addition to the desired two-photon CP emission, first-order Bogoliubov quasiparticle (BQP) emission can also occur. These frequencies are never emitted by embedding the entire device within an vertical optical cavity similar to the asymmetric ones used in Vertical-Cavity Surface-Emitting Laser (VCSEL) - a high-Q resonator formed by top distributed Bragg reflector (DBR) and a metallic bottom mirror [70,71]. This cavity can be designed to block/suppress unwanted first-order BQP emission by engineering the optical density of states, while simultaneously enhancing two-photon CP emission into selected cavity modes in our device.

From Ginzburg-Landau theory [72] a uniform supercurrent in a loop corresponds to quantized order parameter phase distribution $\oint \boldsymbol{\nabla}\Theta(\boldsymbol{r})\overrightarrow{dl} = l_s$, where the integral is over a closed loop within the superconducting ring, $\Theta(\boldsymbol{r})$ is the phase of the order parameter, and $l_s$ is the winding number. By additionally applying voltage between the superconducting electrodes, the BCS condensates that are introduced into the CB in the *n*-side and the VB in the *p*-side undergo recombination.

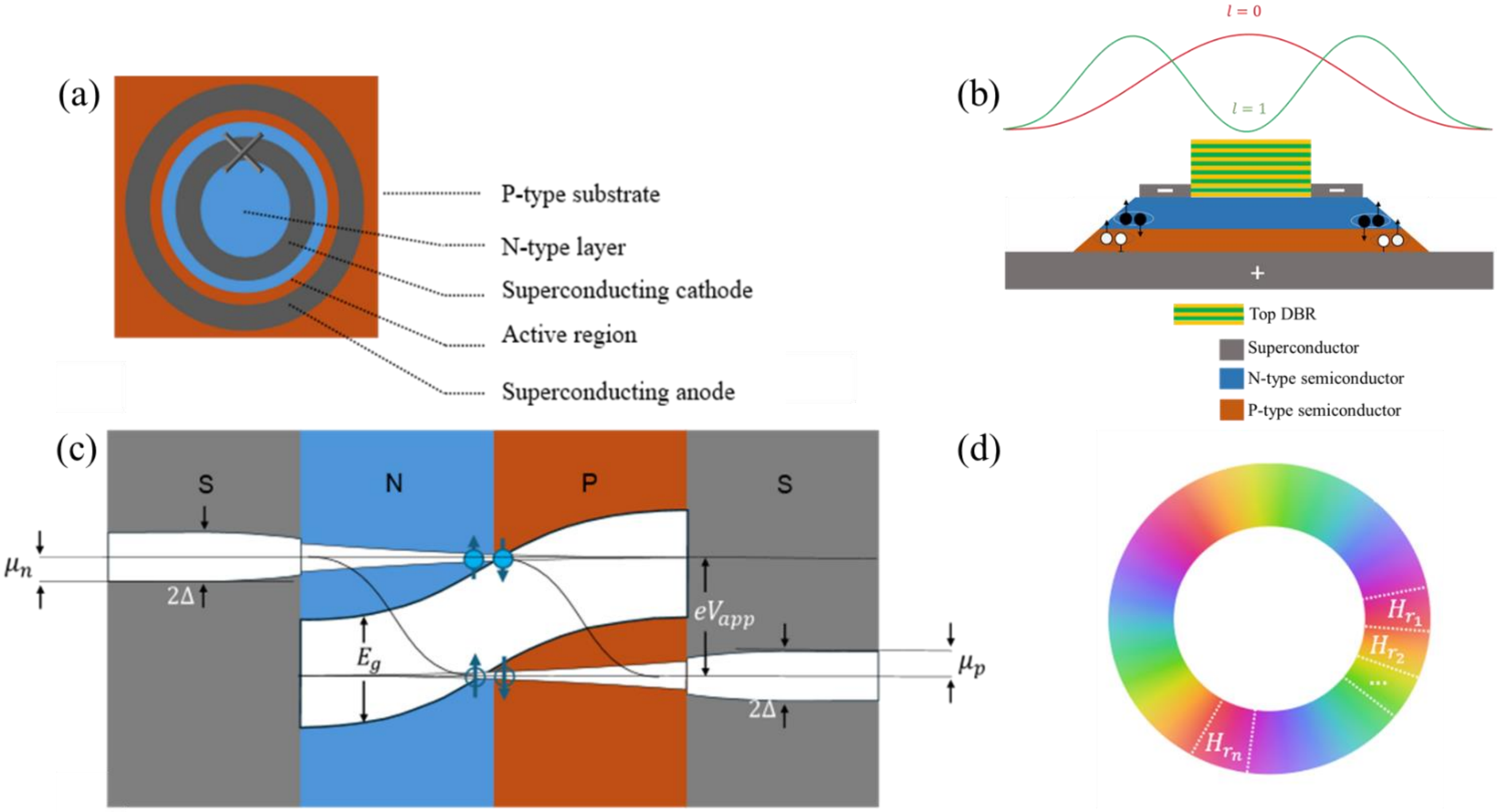

FIG. 1 (a) Top view of the proposed structure with legend describing the different layers. (b) Side view of the structure with cavity to eliminate the BQP one-photon emission, with the plus (minus) signs describing positive (negative) bias voltage. Here bottom superconductor will act as bottom DBR. (c) A qualitative band diagram of the S-N-P-S layers, depicting the semiconductor bandgap, the superconducting gap in the superconducting region, and the superconducting gap penetrating the semiconductor region via the superconducting proximity effect. Quasi Fermi levels are plotted by dashed lines, with the offset between them set by the voltage applied to the junction. (d) Schematic azimuthal phase profile of the BCS order parameter in the ring-shaped cathode superconductor for winding number $l$ = 3. The color represents to the convention of complex phase representation from 0 to 2π (repeating three times for $l$ = 3). Segmentation of the Hamiltonian ($H_{r_n}$) for the theoretical treatment is schematically shown on top, with each segment having approximately a constant phase. Color corresponds to the convention of complex phase representation.

To treat the interaction within the semiconductor we use the BCS theory, with a spatially varying BCS order parameter corresponding to the quantized Ginzburg-Landau order parameter -phase [73]:

$$\Delta(\boldsymbol{r}) = |\Delta(r)|e^{il_s\phi}, \tag{1}$$

where $|\Delta(r)|$ is the magnitude of the BCS gap and the phase is deterministic for the entire condensate. Optical modes carrying OAM are typically described using the Laguerre–Gaussian (LG) basis [60,74,75], whose transverse field profile is characterized by a azimuthal phase factor $e^{il\phi}$. Modes with different $l$ values form an orthonormal basis in the photonic Hilbert space, enabling high-dimensional quantum encoding. In two-photon emission processes, the quantum state of the emitted light is represented by a density matrix $\rho_{l_1,l_2;l_1{}',l_2{}'}$, whose off-diagonal elements encode coherence between different OAM pairs. It directly captures entanglement and coherence properties of multimode photon pairs. The AC Josephson effect, between the top and bottom superconductor, plays a key role in linking the electronic and photonic phases. A voltage bias creates a well-defined phase evolution $d\theta/dt$ = 2eV/$\hbar$, which determines the oscillation frequency of the emitted electromagnetic field. Thus, the spatial phase of the superconducting condensate is imprinted onto the photonic state, providing the physical pathway through which the spatial phase of the order parameter transfers to the emitted OAM-carrying photons.

The radiative recombination of electrons (with total momentum and polarization; $J + \sigma$) and holes (total momentum; $-J$) in the active region is described by the following light-matter Hamiltonian-density in the interaction picture [76]:

$$\mathcal{H}_I(\boldsymbol{r},\boldsymbol{t}) = \sum_{\boldsymbol{k},\boldsymbol{k}',\boldsymbol{q},\sigma} B_\sigma b_{-\boldsymbol{k}',-J}(t) c_{\boldsymbol{k},J+\sigma}(t) a^\dagger_{\boldsymbol{q},\sigma} e^{i(\boldsymbol{k}-\boldsymbol{k}'-\boldsymbol{q})\cdot\boldsymbol{r}} + H.C., \tag{2}$$

where $b^\dagger_{\boldsymbol{k}',\boldsymbol{J}}(t)\left(b_{\boldsymbol{k}',\boldsymbol{J}}(t)\right), c^\dagger_{\boldsymbol{k},\boldsymbol{J}+\boldsymbol{\sigma}}(t)\left(c_{\boldsymbol{k},\boldsymbol{J}+\boldsymbol{\sigma}}(t)\right), a^\dagger_{\boldsymbol{q},\boldsymbol{\sigma}}(a_{\boldsymbol{q},\boldsymbol{\sigma}})$ are the creation (annihilation) operator of the holes, electrons, and photons, respectively with wavevectors $\boldsymbol{k}'$, $\boldsymbol{k}$, $\boldsymbol{q}$; $J$ is the total spin-orbit-coupled angular momentum in the VB, $\sigma$ is the photon polarization and $B_\sigma$ is the polarization-dependent interaction strength.

To account for the long-range coherence properties of the BCS state, namely, the spatial dependence of the order parameter Eq. (1) over the extent of the structure, we integrate locally over space in the first step yielding momentum conservation on the scale of a single CP, followed by integration over the extent of the structure to achieve OAM conservation. To carry out the first step, we divide the active region into small segments labeled by $n$ (Fig. 1 (d)), centered around $\boldsymbol{r}_n$ and chosen so that each segment is larger than the size of a CP but smaller than the length scale of a significant change in either the order parameter phase or the wavelength of the electromagnetic field. Carrying out the small-scale integration yields:

$$H_I(\boldsymbol{r}_n,t) = \int_n d^3\boldsymbol{r}\mathcal{H}_I(\boldsymbol{r},\boldsymbol{t}) = e^{-i\boldsymbol{q}\cdot\boldsymbol{r}_n} \sum_{\boldsymbol{k},\boldsymbol{q},\sigma} B_\sigma b_{-\boldsymbol{k},-J}(t) c_{\boldsymbol{k},J+\sigma}(t) a^\dagger_{\boldsymbol{q},\sigma} + H.C., \tag{3}$$

where $H_I(\boldsymbol{r}_n,t)$ is part of the Hamiltonian for the segment around $\boldsymbol{r}_n$. The initial wavefunction is composed of a BCS state in the CB, a BCS state in the VB and a vacuum photonic state: $|\psi_0(r)> = |\psi^n_{BCS}(r)>\otimes|\psi^p_{BCS}(r)>\otimes|\psi_{ph}>$, where $|\psi_{ph}\rangle$ is the vacuum photon state ($|0_{ph}\rangle$). To obtain the state of the two-photon emission we expand the initial state to second order in perturbation theory:

$$\left|\psi^{(2)}_t(r_n)\right\rangle = \sum_{r_n} \frac{1}{\hbar^2}\int_{-\infty}^{t} dt_2 \int_{-\infty}^{t_2} dt_1\, H_I(r_n,t_2) H_I(r_n,t_1)|\psi_i(r_n)\rangle, \tag{4}$$

To analyze the two-photon state in modes of OAM we evaluate the two-photon density-matrix given by [77]:

$$\rho_{\alpha,\beta,\gamma,\delta}(\omega_\mu,\omega_\nu)=\lim_{t\to\infty}\frac{1}{t}\left\langle\psi_t^{(2)}\middle|a^\dagger_{m_\alpha,l_\alpha,\sigma_\alpha,\omega_\mu}a^\dagger_{m_\beta,l_\beta,\sigma_\beta,\omega_\nu}a_{m_\gamma,l_\gamma,\sigma_\gamma,\omega_\nu}a_{m_\delta,l_\delta,\sigma_\delta,\omega_\mu}\middle|\psi_t^{(2)}\right\rangle, \tag{5}$$

where the operator $a^\dagger_{m,l,\sigma,\omega}$ creates a photon in a Laguerre-Gaussian (LG) mode with OAM quantum number $l$, SAM quantum number $\sigma$, number of nonaxial radial nodes $m$, and energy $\hbar\omega$. From Eq. (2, 4, 5), after converting the sum over space to integration over space (given that the transverse phase $\boldsymbol{q}\cdot\boldsymbol{r}_n$ is slowly varying on the length scale of a single segment), we obtain up to an overall normalization constant:

$$\rho_{\alpha,\beta,\gamma,\delta}(\omega_\mu,\omega_\nu)=\lim_{t\to\infty}\frac{1}{t}\int_{-\infty}^{t}dt_3\int_{-\infty}^{t_3}dt_4\int_{-\infty}^{t}dt_2\int_{-\infty}^{t_2}dt_1\left\{\frac{1}{\hbar^4}\sum_{\{\boldsymbol{k},\boldsymbol{q},\sigma,J\}_{1..4}}B^*_{\sigma_1}B^*_{\sigma_2}B_{\sigma_3}B_{\sigma_4}\times\right.$$

$$\left.e^{i(-\omega_{q_1}t_1-\omega_{q_2}t_2+\omega_{q_3}t_3+\omega_{q_4}t_4)}\int d^3r'\int d^3r\, e^{i(\boldsymbol{q}_1+\boldsymbol{q}_2)\cdot\boldsymbol{r}}e^{-i(\boldsymbol{q}_3+\boldsymbol{q}_4)\cdot\boldsymbol{r}'}Q_n(r,r')Q_p(r,r')Q_{ph}\right\}, \tag{6}$$

where:

$$Q_n(r,r') \stackrel{\text{def}}{=} \langle BCS_n(r')|c^\dagger_{\boldsymbol{k}_1,J_1+\sigma_1}(t_1)c^\dagger_{\boldsymbol{k}_2,J_2+\sigma_2}(t_2)c_{\boldsymbol{k}_3,J_3+\sigma_3}(t_3)c_{\boldsymbol{k}_4,J_4+\sigma_4}(t_4)|BCS_n(r)\rangle, \tag{a.}$$

$$Q_p(r,r') \stackrel{\text{def}}{=} \left\langle BCS_p(r')\middle|b^\dagger_{-\boldsymbol{k}_1,-J_1}(t_1)b^\dagger_{-\boldsymbol{k}_2,-J_2}(t_2)b_{-\boldsymbol{k}_3,-J_3}(t_3)b_{-\boldsymbol{k}_4,-J_4}(t_4)\middle|BCS_p(r)\right\rangle, \tag{b.}$$

$$Q_{ph} \stackrel{\text{def}}{=} \left\langle 0_{ph}\middle|a^\dagger_{\boldsymbol{q}_1,\sigma_1}a^\dagger_{\boldsymbol{q}_2,\sigma_2}\hat{\rho}_{\alpha,\beta,\gamma,\delta}(\omega_\mu,\omega_\nu)a_{\boldsymbol{q}_3,\sigma_3}a_{\boldsymbol{q}_4,\sigma_4}\middle|0_{ph}\right\rangle, \tag{c.}$$

with $\hat{\rho}_{\alpha,\beta,\gamma,\delta}(\omega_\mu,\omega_\nu) \stackrel{\text{def}}{=} a^\dagger_{m_\alpha,l_\alpha,\sigma_\alpha,\omega_\mu}a^\dagger_{m_\beta,l_\beta,\sigma_\beta,\omega_\nu}a_{m_\gamma,l_\gamma,\sigma_\gamma,\omega_\nu}a_{m_\delta,l_\delta,\sigma_\delta,\omega_\mu}$. Since the BCS state is not an eigenstate of $c^\dagger_k c_k$, $b^\dagger_k b_k$, a transformation to the Bogoliubov operators Eq. (7) is needed in order to calculate the electron (hole) term in Eq. 6 (a-b).

$$c^\dagger_{k,\sigma}(t)=e^{\frac{i\tilde{\mu}_n}{\hbar}t}\left(u_k e^{\frac{iE_k}{\hbar}t}\gamma^\dagger_{k,\sigma}+s_\sigma v^*_k e^{-\frac{iE_k}{\hbar}t}\,\gamma_{-k,-\sigma}\right) \tag{7a}$$

$$c_{k,\sigma}(t)=e^{-\frac{i\tilde{\mu}_n}{\hbar}t}\left(u^*_k e^{-\frac{iE_k}{\hbar}t}\gamma_{k,\sigma}+s_\sigma v_k e^{\frac{iE_k}{\hbar}t}\,\gamma^\dagger_{-k,-\sigma}\right) \tag{7b}$$

as was performed in [53]. For the photonic term Eq. (6c), the OAM mode operators are spanned by planewave operators: $a^{\dagger}_{m,l,\sigma,\omega} = \sum_{\boldsymbol{q}} L_{m,l,\omega}(\boldsymbol{q}) a^{\dagger}_{\boldsymbol{q},\sigma}$, where: $L_{m,l,\omega}(\boldsymbol{q}) = \int d^3\boldsymbol{r} u_{m,l,\omega}(\boldsymbol{r}) e^{-i\boldsymbol{q}\cdot\boldsymbol{r}}$ is a weight equal to the Fourier transform of the electric field of the mode – $u_{m,l,\omega}(\boldsymbol{r})$. In our model, both the n-type and p-type regions are proximitized by superconductors, forming a continuous condensate across the p-n junction. This ensures a globally coherent superconducting order parameter and a deterministic relative phase across the structure. The time-dependent behavior of the emitted photon state in our calculations arises from the AC Josephson effect [78], driven by the difference between the quasi-Fermi levels $\mu_n$ and $\mu_p$, which is voltage bias and leads to a deterministic phase evolution of the density matrix. The resulting density matrix acquires a time-dependent phase factor $e^{i(\mu_n - \mu_p)t/\hbar}$, which directly sets the frequency of the emitted photons.

This is explicitly seen in the density matrix calculation where a phase term of the form $e^{\frac{i}{\hbar}(\tilde{\mu_n} - \tilde{\mu_p})\,(t_1+t_2-t_3-t_4)}$ emerges in the matrix element as a consequence of AC Josephson effect:

$$\langle\psi_0(r')|H_I(r',t_1)H_I(r',t_2)\rho\hat{}H(r,t_3)H_I(r,t_4)\,|\psi_0(r)\rangle$$

$$= \sum_{\{k,q,\sigma,J\}1..4} B^*_{\sigma_1} B^*_{\sigma_2} B_{\sigma_3} B_{\sigma_4} e^{i(-\omega_{q_1}t_1 - \omega_{q_2}t_2 + \omega_{q_3}t_3 + \omega_{q_4}t_4)} e^{i(q_1+q_2)\cdot r'} e^{-i(q_3+q_4)\cdot r}\ e^{\frac{i}{\hbar}(\tilde{\mu_n} - \tilde{\mu_p})\,(t_1+t_2-t_3-t_4)}$$

$$\times \left(Q^e_{cp}\,\delta^n_{1,-2}\delta^n_{3,-4} s_{\sigma_1+J_1} s_{\sigma_3+J_3} + Q^n_{e,1}\delta^n_{1,4}\,\delta^n_{2,3} - Q^n_{e,2}\delta^n_{1,3}\ \delta^n_{2,4}\right) \times \left(Q^e_h\,\delta^p_{1,-2}\delta^p_{3,-4} s_{J_1} s_{J_3} + Q^n_{h,1}\delta^p_{1,4}\,\delta^p_{2,3} - Q^n_{h,2}\delta^p_{1,3}\ \delta^p_{2,4}\right)$$

$$\times \Big\{L_{\alpha\mu}(q_1)L_{\beta\nu}(q_2)L^*_{\gamma\nu}(q_3)L^*_{(\delta\mu)}(q_4)\delta_{\sigma_1\sigma_\alpha}\delta_{\sigma_2\sigma_\beta}\,\delta_{\sigma_3\sigma_\gamma}\,\delta_{\sigma_4\sigma_\delta} + (\sigma_1 \leftrightarrow \sigma_2) + (\sigma_3 \leftrightarrow \sigma_4) + (\sigma_1 \leftrightarrow \sigma_2)\times(\sigma_3 \leftrightarrow \sigma_4)\Big\},$$

(8)

where $|\psi_0(r)> = |\psi^n_{BCS}(r)> \otimes |\psi^p_{BCS}(r)> \otimes |\psi_{ph}>$

The Fermi level difference term arises from the superconductor part of the calculation and is precisely matched by the emitted photons energies, ensuring energy conservation in the recombination process. Consequently, the oscillation frequency of the electromagnetic field in our model is set by Fermi energy difference on both sides, fully consistent with the AC Josephson effect, and confirms deterministic phase evolution in the density matrix.

Performing integration over time and over all space in Eq. (6) for a quantum-well active region, and starting with $l_s = 0$ for the anode condensate but a nonzero $l_s$ for the cathode condensate, we obtain for the CP-related terms of Eq. (6) (up to an overall normalization constant):

$$\rho^{cp}_{\alpha,\beta,\gamma,\delta}(\omega_\mu, \omega_\nu) = \left|B_{\sigma_\alpha} B_{\sigma_\gamma}\right|^2 \delta_{\sigma_\alpha,-\sigma_\beta}\delta_{\sigma_\gamma,-\sigma_\delta}\delta_{l_\alpha+l_\beta,l_s}\delta_{l_\gamma+l_\delta,l_s}\mathrm{K}^{cp}_{\alpha,\beta,\gamma,\delta}S^{cp}(\omega_\mu, \omega_\nu, T), \tag{9}$$

where $\mathrm{K}^{cp}_{\alpha,\beta,\gamma,\delta}$ is the overlap of transverse field profiles in the active region. It is given by $\mathrm{K}^{cp}_{\alpha,\beta,\gamma,\delta} = \kappa_{m_\alpha m_\beta\, l_\alpha l_\beta}\kappa_{m_\gamma m_\delta\, l_\gamma l_\delta}$, where:

$$\kappa_{m_\alpha m_\beta\, l_\alpha l_\beta} = \int_{\substack{active\\ region}} u_{m_\alpha,l_\alpha}(r) u_{m_\beta,l_\beta}(r) r dr, \tag{10}$$

with $u_{m,l}(r)$ the transverse part of $u_{m,l,\omega}(\boldsymbol{r})$ in the emitting region plane. The term $\delta_{l_\alpha+l_\beta,l_s}\delta_{l_\gamma+l_\delta,l_s}$ is the manifestation of OAM conservation resulting from the integral over the spatial extent of the structure, where the phases of each of the modes added to the phase of the superconducting order parameter are integrated over. The function $S^{cp}(\omega_\mu, \omega_\nu, T)$ is proportional to the two-photon rate of the CP-related emission, and is given by:

$$S^{cp}(\omega_\mu, \omega_\nu, T) = \delta(\hbar\omega_\mu + \hbar\omega_\nu - 2U_{os})\left|\sum_k |v^n_k u^n_k v^p_k u^p_k|\right.$$

$$\times\left(\frac{\bar{f}(E_k^n)\bar{f}(E_k^p)}{d_{\omega_\mu}-\frac{E_k^n}{\hbar}-\frac{E_k^p}{\hbar}+\frac{i}{\tau}}+\frac{\bar{f}(E_k^n)f(E_k^p)}{d_{\omega_\mu}-\frac{E_k^n}{\hbar}+\frac{E_k^p}{\hbar}+\frac{i}{\tau}}\right.$$

$$\left.\left.+\frac{f(E_k^n)\bar{f}(E_k^p)}{d_{\omega_\mu}+\frac{E_k^n}{\hbar}-\frac{E_k^p}{\hbar}+\frac{i}{\tau}}+\frac{f(E_k^n)f(E_k^p)}{d_{\omega_\mu}+\frac{E_k^n}{\hbar}+\frac{E_k^p}{\hbar}+\frac{i}{\tau}}\right)+\left(d_{\omega_\mu}\leftrightarrow d_{\omega_\nu}\right)\right|^2, \quad (11)$$

where $u_k^i(v_k) = \left[1+(-){\xi_k^i}^2/{E_k^i}^2\right]/2$ , $\xi_k^i = \frac{\hbar^2 k^2}{2m^i} - \mu_F^i$ , $E_k^i = \sqrt{|\Delta|^2 + {\xi_k^i}^2}$ with $i$ denoting either $p$ for the VB or $n$ for the CB, $\mu_F^i$ the quasi Fermi-level measured from the bottom of the band, $m^i$ the effective mass of the quasi-particles in either of the bands, $f$ the Fermi-Dirac distribution function, $\bar{f} \stackrel{\text{def}}{=} 1 - f$, $U_{os} \stackrel{\text{def}}{=} E_{gap} + |\mu_F^n| + |\mu_F^p|$ is the energy offset between the quasi Fermi-levels of the $p$ and $n$ side of the semiconductor and $d_\omega = \omega - \frac{U_{os}}{\hbar}$ is the detuning from $U_{os}$ in units of angular frequency. Equation (11) encapsulates the spectral dependence of the two-photon rate (Fig. 2(a)), which exhibits resonances wherever the virtual state coincides with one of the real states of the system. In real systems this resonance behavior is broadened as is represented by the dephasing term $i/\tau$ in the denominator [53], where $\tau$ is typically about a few hundreds of femto-seconds and is related to scattering mechanisms of semiconductor electrons [79].

In the expansion of Eq. (6a) and Eq. (6b), after performing the Bogoliubov transformation and applying Wick's theorem, the calculation yields a total set of 16 terms. Removing the four terms with an unequal number of creation and annihilation operators and using Wick's theorem, we are left with twelve terms, four terms correspond to CP recombination processes and are used in deriving Eq. (9), while the remaining eight terms correspond to BQP recombination processes. The latter contribute to the emission in a random manner, which if considered alone would result in a mixed (diagonal) density matrix due to the incoherent nature of the BQP [80]. A comparison between the CP-related rate – $R_{cp}$ and the BQP-related rate – $R_{bqp}$ as a function of photon-energy detuning and temperature is presented in Fig. 2, where the rates are normalized to the maximum of the CP-related rate and computed for a GaAs-AlGaAs based quantum-well with Nb contacts, normal electron and hole density of $10^{12}[cm^{-2}]$ and broadening given by $\tau = 1000\ fs$. Recent experiments have reported two-photon spectral features and pair correlations consistent with CP–mediated emission in proximitized SLEDs [35,57]. This efficiency depends on the competition between CP emission and quasiparticle-related or nonradiative loss channels. Strategies to reduce BQP contribution include operation at $T << T_c$ to suppress thermal quasiparticles and using the cavity to reduce the effect of BQP one photon process. For a single superconducting contact SLED device, the emission rate for the two-photon vs the single photon have been found comparable [35,48]. The two-photon rate scales quadratically with emitting perimeter ($\sim L^2$) due to constructive interference across the macroscopic condensate, while incoherent BQP processes scale linearly ($\sim L$). As a result, in our double superconducting contact SLED, the desired CP emission rate can scale more favorably with device size than competing quasiparticle loss channels, providing an additional route to improving the effective two-photon emission efficiency. These approaches, taken together, are expected to substantially increase photon pair and reduce the influence of BQPs in practical devices.

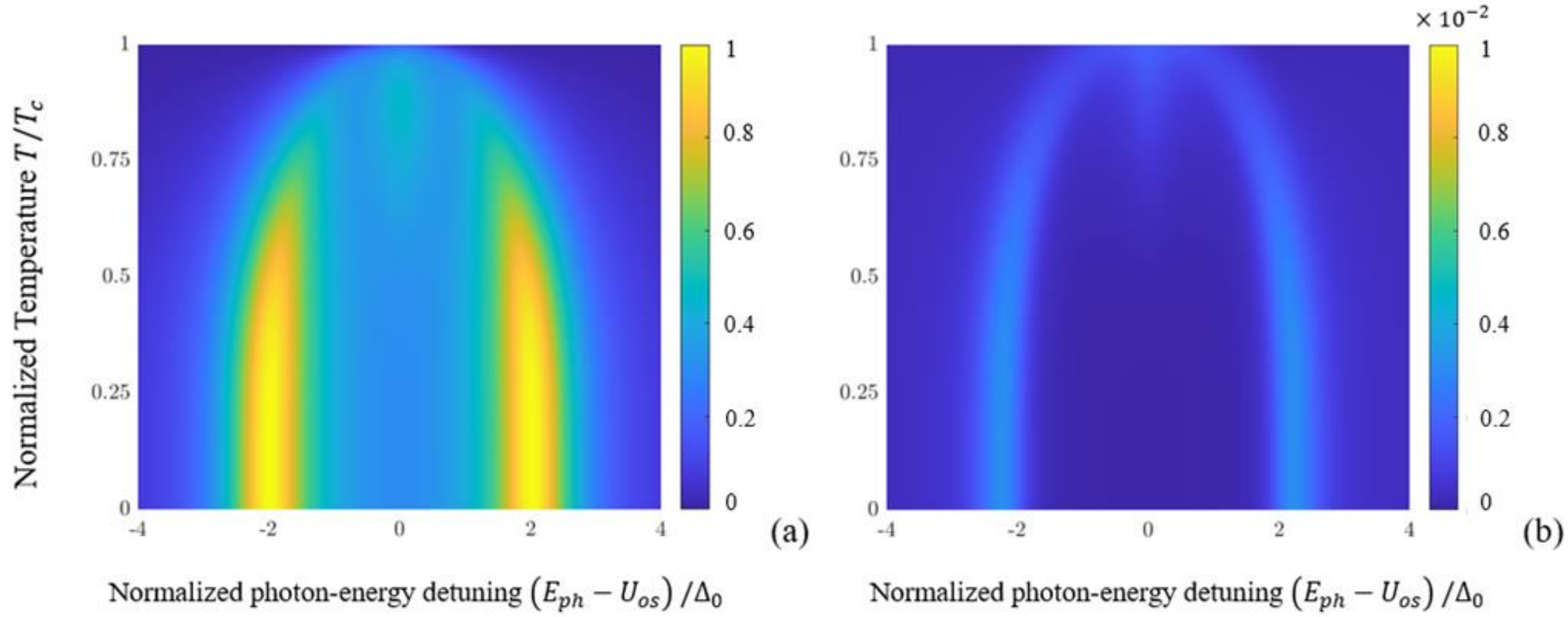

FIG. 2 Two-photon rate as a function of photon-energy detuning and temperature; (a) For CP recombination process, and (b) for second-order BQP recombination processes. The calculation was done with ring-shaped emitting region $L = 10\ \mu m$, and quasi-particle coherence length $L_\phi = 100\ nm$. Both plots are normalized to the maximum of the CP-related emission rate (the maximum of (a)). Color bar shows the two photon rates in each panel. It can be seen that the rate of BQP rate is lower than the CP related two photon emission rates.

The emission rate was computed which is symmetric about zero detuning since photons are emitted in pairs of positive and negative detuning. Since the CP-related processes are coherent over the circumference of the ring $L$, as opposed to BQP-related processes which have a much smaller dephasing length $L_\phi$, the CP rate is enhanced by a factor of $\frac{L}{L_\phi}$ in comparison with the BQP rate. Figure 2 was calculated for a ring circumference of $10\ \mu m$ which is on the scale of superconducting qubits [81], and for $L_\phi$ set to a typical value of $100\ nm$ [82], which makes this enhancement factor equal to $100$. At low temperatures, the emission is enabled by the creation of two BQP in Fig. 2(b) or by a transition through a virtual state containing 2 BQP in the case of Fig. 2(a). Therefore, at low temperatures the peaks are centered at $\pm 2|\Delta|$ detuning due to the increase in the joint density of states in energy offsets of $\pm|\Delta|$ around the superconducting gap in each band (Fig. 1(c)). At temperatures closer to $T_c$, there is the additional possibility of breaking a CP via the creation of one BQP, and the destruction of another thermally excited one. This possibility enables emission of a photon pair around zero detuning, as is evident by the peak around zero detuning at $T > 0.7\,T_c$ both in Fig. 2(a) and (b). As the temperature increases, the emission rate shifts towards zero detuning due to superconducting gap narrowing as described by the equation $\Delta(T) = \Delta_0 \tanh\left(1.74\sqrt{\frac{T_c}{T} - 1}\right)$ [72], and the CP emission rate decreases due to the decrease in CP density which is proportional to $|\Delta(T)|$.

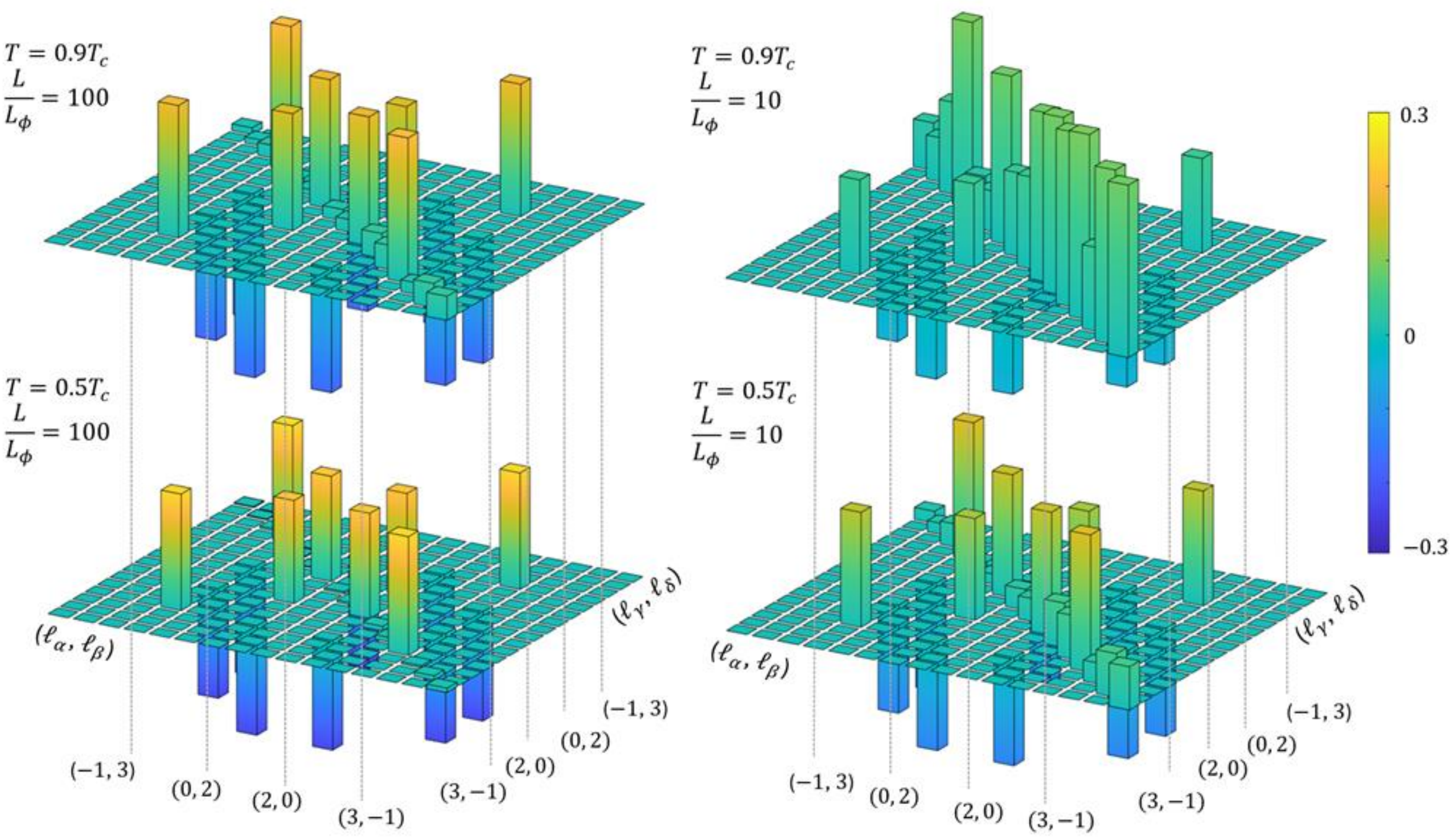

FIG. 3 OAM density matrix $\rho_{l_\alpha l_\beta, l_\gamma l_\delta}$ calculated for the case $l_s = 2$ with maximum detected OAM of $|l| = 3$, $(m)_{\alpha,\beta,\gamma,\delta} = 0$, and energy detuning $\pm 5\Delta_0$. The operating temperature is $T = 0.5\,T_c$ (bottom) and $T = 0.9\,T_c$ (top) and the coherence enhancement factor $L/L_\phi$ is 100 (left) and 10 (right). OAM quantum numbers are shown on the x and y axes. The ticks of each axis go over all possibilities out of (-1,0,2,3) for both OAM quantum number. The representation of the matrices here does not include mode combinations which cannot be populated from CP recombination processes when $l_s = 2$ due to conservation of OAM.

First order BQP radiative recombination has not been considered in our analysis since its bandwidth is narrow in comparison with second-order processes owing to the fact that each photon on its own has to match the offset energy $U_{OS}$. Thus, operating in photon-energy detuning larger than the spectral broadening, first order contribution is prevented by using a cavity selecting only desired photon energy. To demonstrate OAM photon entanglement, we start with a circular current which imparts a phase winding number of $l_s = 2$ on the cathode condensate, while the anode has $l_s = 0$. Thus, the two emitted photons can only populate modes with OAM quantum numbers summing up to two. For simplicity, we consider an apparatus that detects modes with $l = 0, \pm1, \pm2, \pm3$, and we calculate a density matrix of the $l$ quantum numbers, with the arbitrary choice of $(m)_{\alpha,\beta,\gamma,\delta} = 0$ and any choice for the polarization $(\sigma)_{\alpha,\beta,\gamma,\delta}$, as given by Eq. (9). In Fig. 3, the total density matrix for this case, including both CP emission and BQP emission, is presented for photon-energy detuning of $\pm5\Delta_0$ for two different operating temperatures and coherence enhancement factors $L / L_\phi$ . The magnitude of different matrix elements is varied due to the $K_{\alpha,\beta,\gamma,\delta}$ factor which modulates the contribution of each OAM mode pair depending on the geometry of the emitting structure (Eq. 10), as is evident even for low temperature and high coherence enhancement. For temperatures close to $T_c$ the diagonal matrix elements become more prominent than at low temperatures due to the mixing of BQP-related photons. This effect is pronounced for small values of $L / L_\phi$ , while for large $L / L_\phi$ the enhanced emission of the coherent CP-related processes makes these dominant even closer to $T_c$. The total density matrix $\rho_{l_\alpha,l_\beta,l_\gamma,l_\delta}$ was computed as a weighted sum of the pure density matrix given by CP-related photons given up to normalization by: $(\rho^{cp})_{l_\alpha,l_\beta,l_\gamma,l_\delta} = \mathrm{K}_{\alpha,\beta,\gamma,\delta}\delta_{l_\alpha+l_\beta,l_s}\delta_{l_\gamma+l_\delta,l_s}$ and a mixed density matrix given up to normalization by: $(\rho^{bqp})_{l_\alpha,l_\beta,l_\gamma,l_\delta} = \mathrm{K}_{\alpha,\beta,\gamma,\delta}\delta_{l_\alpha,l_\gamma}\delta_{l_\beta,l_\delta}$ with relative weights given by the corresponding rates (up to normalization):

$$\rho_{l_\alpha,l_\beta,l_\gamma,l_\delta} = R_{bqp}\frac{\rho^{bqp}}{tr(\rho^{bqp})} + R_{cp}\frac{\rho^{cp}}{tr(\rho^{cp})} \tag{12}$$

Bell states are foundational in quantum optics as benchmarks for entanglement, they achieve maximum Bell-inequality violation and enable key protocols like quantum teleportation and dense coding in high-dimensional spaces. A representative maximally entangled state is the photonic Bell state, e.g., (1/√2)(|2, 0⟩ + |0, 2⟩). Such Bell states are widely used as standard benchmarks for entanglement generation and verification in quantum optics. We demonstrate that coherent superposition of superconducting qubit eigenstates are mapped onto a coherent superposition of photon states. We consider an initial qubit superposition state given by:

$$|\psi\rangle = a|\circlearrowright> + b|\circlearrowleft> \tag{13}$$

where $|\circlearrowright>$ and $|\circlearrowleft>$ represents the clockwise and counterclockwise states respectively and the complex coefficients $a$ and $b$ determine the relative amplitude and phase of each component. We show that this superposition in the qubit basis can be faithfully transferred to a superposition in the photon basis—this should manifest clearly in the resulting density matrix. To this end, we substitute the above qubit superposition into the density matrix formulation presented in Eq. (6). In this scenario, instead of the single term $\delta_{l_\alpha+l_\beta,l_s}\delta_{l_\gamma+l_\delta,l_s}$ which corresponds to both emitted photons having a total angular momentum $\hbar l_s$ owing to the initial clockwise state of the qubit current, we now obtain the full expression:

$$\rho^{cp}_{\alpha,\beta,\gamma,\delta}(\omega_\mu,\omega_\nu) = \left|B_{\sigma_\alpha} B_{\sigma_\gamma}\right|^2 \delta_{\sigma_\alpha,-\sigma_\beta}\delta_{\sigma_\gamma,-\sigma_\delta}K^{cp}_{\alpha,\beta,\gamma,\delta}\, S^{cp}\,(\omega_\mu,\omega_\nu,T)[|a|^2\delta_{l_\alpha+l_\beta,l_s}\delta_{l_\gamma+l_\delta,l_s} + ab^*\delta_{l_\alpha+l_\beta,l_s}\delta_{-l_\gamma-l_\delta,l_s} + a^*b\delta_{-l_\alpha-l_\beta,l_s}\delta_{l_\gamma+l_\delta,l_s} + |b|^2\delta_{-l_\alpha-l_\beta,l_s}\delta_{-l_\gamma-l_\delta,l_s}], \tag{14}$$

which corresponds to both emitted photons having a total angular momentum of $\hbar l_s$ in a coherent superposition with corresponding coefficients $a, b$. The structure of the resulting density matrix (Eq. 14) clearly demonstrates that the superposed qubit state $|\psi> = a|\circlearrowright> + b|\circlearrowleft>$, is transferred to the total OAM of the photon pair, with the coefficients $a$ and $b$ controlling the relative contributions of each component. In Fig. 4, we present the complete OAM density matrix for two different choices of the superposition coefficients $a$ and $b$ in the qubit state $|\psi\rangle = a\ |↻\rangle + b\ |↺\rangle$. The magnitude of off-diagonal elements in the density matrix, as shown in Fig. 4, confirms the coherence of the superposition, with the purity and structure of the matrix reflecting the underlying quantum state faithfully. This theoretical analysis not only illustrates the impact of varying phase and amplitude on the photon output but also demonstrates that the phase information encoded in the qubit is faithfully transferred to the photonic system. In particular, the relative phase between |↻> and |↺> manifests in the off-diagonal terms of the density matrix, enabling full reconstruction of the qubit's coherent superposition state through photon state tomography alone.

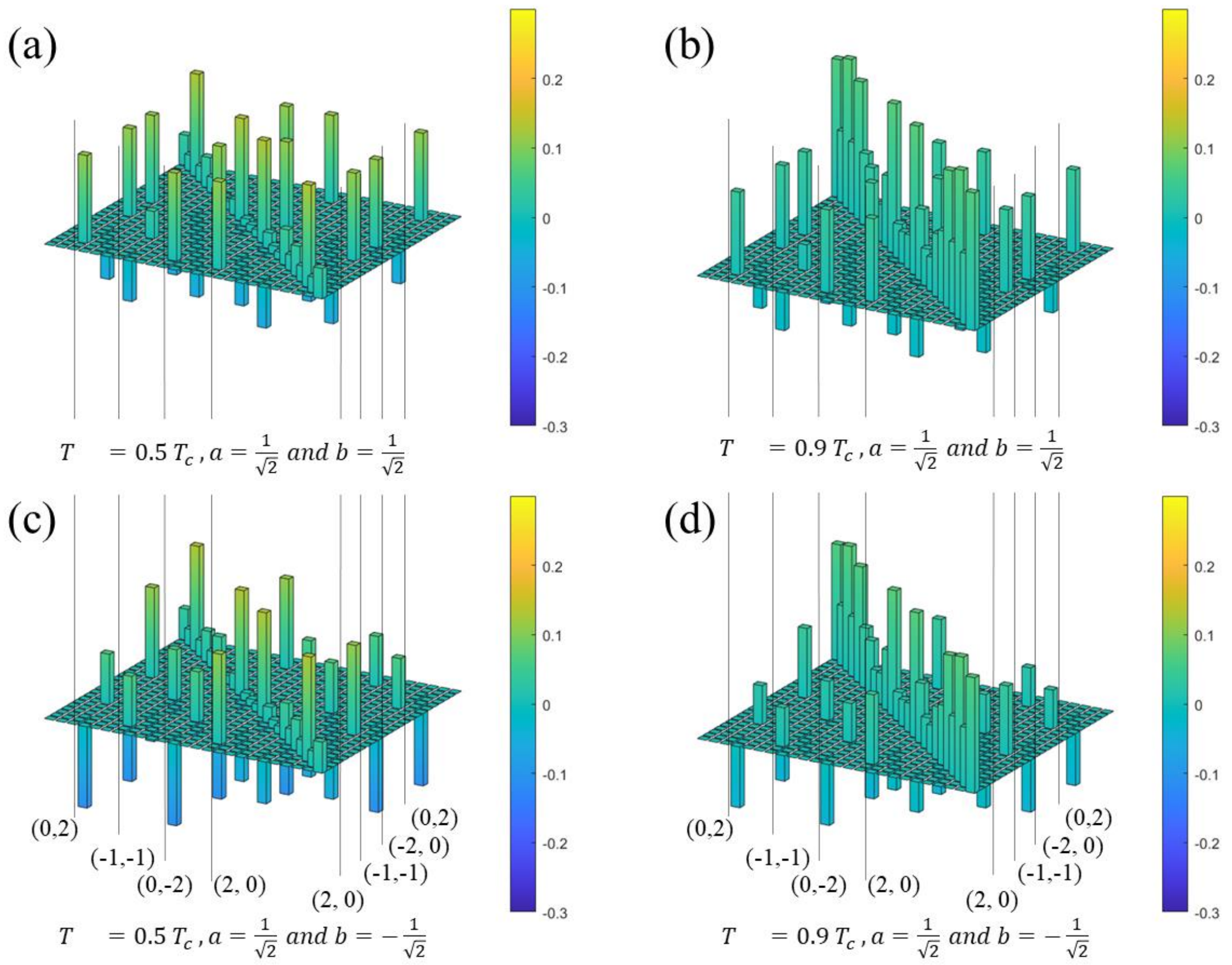

FIG. 4 OAM density matrix $\rho_{l_\alpha l_\beta, l_\gamma l_\delta}$ calculated for the superposition state $|\psi> = \ a|↻> +b|\ ↺>$, with a maximum detected OAM of $|l| = 2$, $(m)_{\alpha,\beta,\gamma,\delta} = 0$, and energy detuning $\pm 5\Delta_0$, shown for two different values of a and b. (a-b) $a\ =\ 1/\sqrt{2}$ and

$b = 1/\sqrt{2}$ at T $= 0.5\ T_c$ and 0.9 $T_c$ respectively; (c-d) $a = 1/\sqrt{2}$ and $b = -1/\sqrt{2}$ at T $= 0.5\ T_c$ and 0.9 $T_c$ respectively. The coherence enhancement factor $L/L_\phi$ is 100. The x- and y-axes label the OAM quantum numbers, ranging over all combinations of $(-2, -1, 0, 1, 2)$.

Experimental verification of the generated OAM entanglement can be performed using existing and widely adopted experimental techniques. Photons are collected with a high-NA objective and analyzed via standard OAM projection methods: fork holograms/SLMs [55,56], q-plates/spiral phase plates [57], or log-polar mode sorters [58,59]. These achieve >90–95% mode purity and >98% crosstalk suppression for $|l| \leq 5$, sufficient for our ±1, ±2 OAM values. Coincidence detection with single photon avalanche diode (SPADs) or superconducting nanowire single photon detectors (SNSPDs) reveals the expected correlations and density-matrix elements (as in Figs. 3–4). The same setup has routinely verified OAM entanglement from spontaneous parametric down-conversion sources [55,56].

Figure 4 shows the emergence of two-photon OAM states from the superconducting-semiconducting hybrid structure. Quantum information transfer additionally requires preservation of quantum coherence. Specifically, an effective quantum interface must preserve quantum coherence, allowing a coherent superposition in the superconducting segment to be faithfully mapped onto a corresponding superposition of photonic states. We quantitatively assess the transfer of phase coherence by evaluating the fidelity ($F$) of the emitted two-photon state with respect to the corresponding ideal photonic superposition. By evaluating the $F$ (where density matrix is the full emitted density matrix from Eq. (12), including realistic CP and BQP contributions), we quantify how closely our device approaches this ideal, under varying temperature and coherence enhancement, directly building on the general superposition results in Fig. 4 and validating the scheme's potential for quantum information transfer. We calculated the $F$ between the photonic state achieved by the emission as described by the density matrix $\rho$ in Eq. (12) for the case of $l_s = 1$ with detected OAM modes of $l = 0, 1$, and the two-photon Bell-state

$\left|\psi_{ph}\right\rangle = \frac{1}{\sqrt{2}}(|l_1 = 1, l_2 = 0\rangle + |l_1 = 0, l_2 = 1\rangle)$. It is given by [83] $F = \left(Tr\left(\sqrt{\sqrt{\rho^i}\rho\sqrt{\rho^i}}\right)\right)^2$ where $\rho^i \overset{\text{def}}{=} |\psi\rangle\langle\psi|$, and presented in Fig. 5. The density matrix $\rho$ was calculated for the same setting as Fig. 3 except for $l_s, l_{max}$.

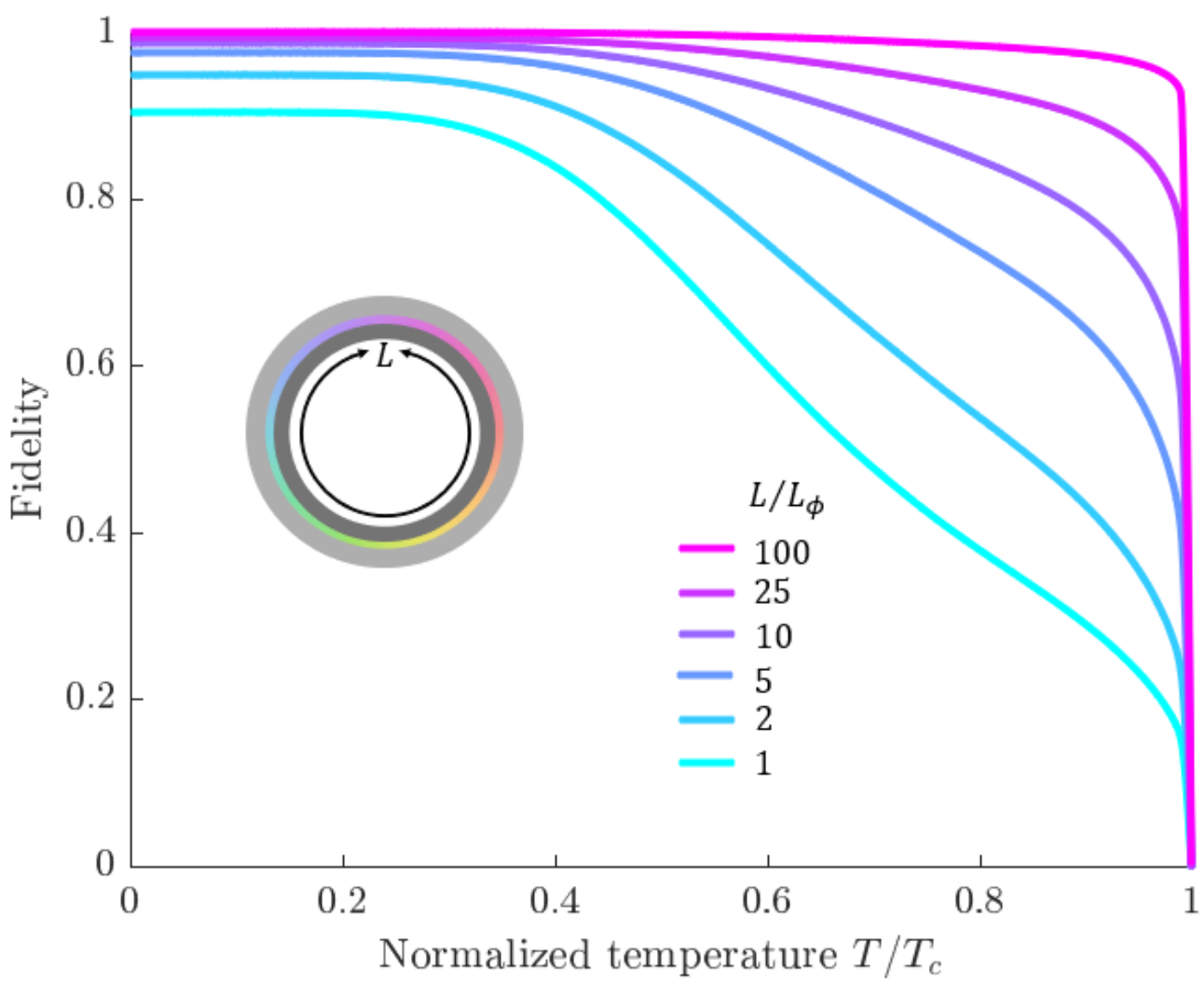

FIG. 5 Fidelity between the emitted two-photon state and the two-photon Bell-state $\left|\psi_{ph}\right\rangle = \frac{1}{\sqrt{2}}(|l_1 = 1, l_2 = 0\rangle + |l_1 = 0, l_2 = 1\rangle)$ as a function of normalized temperature for different enhancement factors $\frac{L}{L_\phi}$, (inset – the colored area between the two contact rings marks the emitting region which has a circumference $L$). The calculation was done for winding number $l_s = 1$, detected OAM modes of $l = 0, 1$, and for the same setting as Fig. 3 for the rest of the parameters.

The state purity is improving with a greater enhancement factor of $L/L_\phi$ which makes the CP-related emission more pronounced than the BQP-related emission. At temperatures greater than about $0.2T_c$ the fidelity starts to deteriorate gradually, as is visible for small values of $L/L_\phi$ , due to the increased population of thermally excited BQP. As the temperature approaches $T_c$, the narrowing of the BCS gap becomes the main mechanism leading to a decline in fidelity as it both reduces the CP density and increases the probability of thermal excitations.

In conclusion, we have theoretically demonstrated a novel mechanism for generating OAM-entangled photon pairs directly from the macroscopic phase of a superconducting persistent-current qubit as well as transferring the superconducting qubit state into a photonic one. By combining Ginzburg-Landau and BCS theories, we showed that radiative recombination of CP in a ring-shaped superconductor-semiconductor-superconductor structure inherits the quantized phase information of the supercurrent, faithfully mapping the qubit states |↻⟩ and |↺⟩ and their coherent superpositions onto photon pairs with total OAM $\pm l_s\hbar$. This approach offers several important advantages over conventional polarization-entangled SLEDs: it enables direct electrical control and transfer of superconducting qubit information into high-dimensional photonic OAM modes, benefits from macroscopic coherence enhancement that sustains high-fidelity entanglement even at moderate temperatures, is fully compatible with standard flux-qubit manipulation and readout. Moreover, the generated entanglement can be verified using established single-photon OAM tomography techniques. Our results establish a new bridge between superconducting quantum computers and photonic quantum communication in an unbounded Hilbert space, opening promising avenues for hybrid quantum networks, high-capacity quantum communication, and scalable quantum information processing.

**Disclosures**

The authors declare no conflicts of interest.

**Data availability statement**

No data were generated or analyzed in the presented research.